\documentclass[pdflatex,sn-mathphys-ay,iicol]{sn-jnl}

\usepackage{graphicx}%
\usepackage{multirow}%
\usepackage{amsmath,amssymb,amsfonts}%
\usepackage{amsthm}%
\usepackage{mathrsfs}%
\usepackage[title]{appendix}%
\usepackage{xcolor}%
\usepackage{textcomp}%
\usepackage{manyfoot}%
\usepackage{booktabs}%
\usepackage{algorithm}%
\usepackage{algorithmicx}%
\usepackage{algpseudocode}%
\usepackage{listings}%

\usepackage{lipsum}
\usepackage{ulem}

\newcommand\iso[2]{\ensuremath{^{#2}\mathrm{#1}}}
\newcommand\ose{\ensuremath{^{16}\mathrm{O}/^{18}\mathrm{O}}}
\newcommand{\msun}{M_\odot}

\begin{document}

\title[Localized $^{18}$O production in white dwarf mergers]{Localized $^{18}$O production in white dwarf mergers}


\author*[1,2]{\fnm{Alexander} \sur{Holas}}\email{alexander.holas@mailbox.org}
\equalcont{These authors contributed equally to this work.}

\author[1,3]{\fnm{Veronica} \sur{Agaeva}}
\equalcont{These authors contributed equally to this work.}

\author[1,3,4]{\fnm{Friedrich K.} \sur{R\"opke}}

\author[2,5]{\fnm{Samuel W.} \sur{Jones}}

\author[1]{\fnm{Javier} \sur{Mor\'an-Fraile}}

\author[1,4]{\fnm{Marco} \sur{Vetter}}

\author[6]{\fnm{R\"{u}diger} \sur{Pakmor}}

\author[7,8,1]{\fnm{Philipp} \sur{Podsiadlowski}}

\affil*[1]{\orgname{Heidelberger Institut f\"ur Theoretische Studien}, \orgaddress{\street{Schloss-Wolfsbrunnenweg 35}, \postcode{69115} \city{Heidelberg}, \country{Germany}}}

\affil[2]{\orgname{The NuGrid collaboration}}

\affil[3]{\orgdiv{Institut f\"ur Theoretische Astrophysik}, \orgname{Zentrum f\"ur Astronomie der Universit\"at Heidelberg}, \orgaddress{\street{Philosophenweg 12}, \postcode{69120} \city{Heidelberg}, \country{Germany}}}

\affil[4]{\orgdiv{Astronomisches Recheninstitut}, \orgname{Zentrum f\"ur Astronomie der Universit\"at Heidelberg}, \orgaddress{\street{M{\"o}nchhofstra{\ss}e 12--14}, \postcode{69120} \city{Heidelberg}, \country{Germany}}}

\affil[5]{\orgdiv{Theoretical Division}, \orgname{Los Alamos National Laboratory}, \orgaddress{\street{Los Alamos}, \city{NM} \postcode{87545}, \country{USA}}}

\affil[6]{\orgname{{Max-Planck-Institut f\"ur Astrophysik}, \orgaddress{\street{Karl-Schwarzschild-Str. 1}, \postcode{85748} \city{Garching}, \country{Germany}}}}

\affil[7]{\orgname{London Centre for Stellar Astrophysics}, \orgaddress{\city{London} \postcode{SW8 2LR}}, \country{UK}}

\affil[8]{\orgname{University of Oxford}, \orgaddress{\street{St Edmund Hall}, \city{Oxford} \postcode{OX1 4AR}}, \country{UK}}


\abstract{
The merger of a He white dwarf (WD) and a CO WD is the favored formation channel for R Coronae Borealis (RCB) stars.
These stars exhibit \ose{} ratios that are orders of magnitude lower than the solar value.
However, it is not fully understood whether such low \ose{} ratios can be achieved in WD merger remnants for the predicted lifetime of RCB stars of around $10^4\,\mathrm{years}$.
In this work, we perform detailed nucleosynthesis calculations of a 3D magnetohydrodynamical simulation of a merger of a $0.3\,\msun$ He WD and a $0.6\,\msun$ CO WD for $4000\,\mathrm{s}$ at which point a steady state in temperature and density is reached. 
From this point, we follow several radial zones to study the long-term production of \iso{O}{18} and its variability throughout the burning region.

We find that the asymmetric merger process leaves an imprint on the distribution of the abundances at the end of our hydrodynamic simulation.
During the long-term evolution up to $100\,\mathrm{years}$, we observe \ose{} ratios of order of unity, although the timescale on which \iso{O}{18} is destroyed again is highly location dependent.
Importantly, our calculations suggest that in the outer layers of the burning shell, the dominant production channel is $^{14}\mathrm{C}(\alpha,\gamma)^{18}\mathrm{O}$ instead of the commonly considered $^{14}\mathrm{N}(\alpha,\gamma)^{18}\mathrm{F}(\beta^+)^{18}\mathrm{O}$ reaction, whereby the former can be sustained for longer periods of time.
Furthermore, these outer regions do not reach the conditions necessary for fast $\alpha$-captures in \iso{O}{18} to \iso{Ne}{22}, thus being favorable to maintaining a low \ose{} ratio.
}

\keywords{binaries: close, stars: abundances, stars: white dwarfs}



\maketitle

\section{Introduction}\label{sec:intro}
R Coronae Borealis (RCB) stars \citep{webbink1984a,clayton1996a} are commonly associated with the product of a merger of a He white dwarf (WD) with a CO WD as the favored formation channel \citep{iben1984a,clayton2007a,clayton2012a}.
Although their precise origin is still disputed --- for example, a He shell flash in a post-asymptotic giant branch star has also been suggested --- recent modeling efforts (e.g., \citealt{staff2012a,staff2018a,crawford2020a,shiber2024a}) were successful in matching observed quantities with WD merger models.
A common characteristic of RCB stars is \ose{} ratios which are considerably lower than the solar value of approximately $500$ (e.g., \citealt{clayton2007a,karambelkar2022a,mehla2025a}), with values on the order of unity recently observed\footnote{Recent work by \cite{mehla2025a} suggests that observed RCB stars are classified by \ose{} ratios above $1$, whereas a ratio below unity indicates a hydrogen-deficient carbon star. We exclude this class of stars from present discussions; although related, they require their own dedicated study. See also, e.g., \cite{crawford2024a} and references therein for more details.}.
There are other characteristics of RCB stars such as their photometric variability due to dust formation, their metallicity, and $^{13}\mathrm{C}/^{12}\mathrm{C}$ ratio, but their low \ose{} ratio is the criterion at the center of our work.

During the merger of a CO WD with a He WD, the lower mass He WD donor will become tidally disrupted and form an accretion disk around the CO WD accretor (see, e.g., \citealt{longland2011a,staff2012a,staff2018a,shiber2024a}).
In addition to the accretion disk, the He will form a thin spherical shell around the CO core, where temperatures well above $1\times10^8\,\mathrm{K}$ will be reached.
This so-called shell-of-fire (SoF) is considered to be the main production site of \iso{O}{18} through the reaction $^{14}\mathrm{N}(\alpha,\gamma)^{18}\mathrm{F}(\beta^+)^{18}\mathrm{O}$, where the precise production yields depend on factors such as progenitor metallicity and the mass ratio between the donor and the accretor \citep{staff2012a}.

Here, two main problems arise that make matching models to the observations difficult.
For one, the temperatures commonly found in models of the SoF are too high to maintain \iso{O}{18} for long periods of time as it will undergo fast $\alpha$-capture reactions $^{18}\mathrm{O}(\alpha,\gamma)^{22}\mathrm{Ne}$. In other words, \iso{O}{18} will be destroyed before it can build up in the amount necessary to match observed \ose{} ratios.
Given the expected lifetime of RCB stars of around $10^4\,\mathrm{years}$ \citep{crawford2020a}, this reaction provides a challenge for current models.
The second problem is that dynamic dredge-up of \iso{O}{16} has been found to negatively impact the modeled \ose{} ratios due to an overabundance of \iso{O}{16} \citep{staff2018a,shiber2024a}.

Although current models have made great strides in investigating the one-dimensional (1D) stellar evolution (e.g., \citealt{zhang2014a,lauer2019a,schwab2019a,crawford2020a}) of RCB stars and the merger process with three-dimensional (3D) hydrodynamic simulations (e.g., \citealt{longland2011a,longland2012a,staff2012a,staff2018a,menon2013a,menon2019a,shiber2024a}), the \iso{O}{18} production cycle is still not fully understood in this context.
Particularly hydrodynamic simulations of the merger do not yet capture the full picture of the dynamic process.
There are many studies that model the merger in 1D (e.g., \citealt{menon2013a,schwab2019a,crawford2024a}), but struggle to accurately capture the asymmetric nature of such events.
Furthermore, current 3D hydrodynamic simulations, although successful in investigating mechanisms such as dredge-up, lack detailed modeling of the microphysics involved in this process, such as an in-situ nuclear network to correctly capture the nuclear energy release in the SoF (with the exception of \citealt{longland2011a}\footnote{There are many simulations of WD-WD mergers including a nuclear network, for example \citet{dan2011a,schwab2012a}, but their results have not been examined to the same detail in the context of RCB stars.}), an accurate equation of state, or magnetic fields.
Lastly, recent studies of the SoF \citep{staff2012a,crawford2020a} assume uniform conditions in the SoF, neglecting the dynamic range of conditions found in this region.

In this work, we present a 3D magnetohydrodynamic (MHD) simulation of a merger of a $0.6\,\msun$ CO WD with a $0.3\,\msun$ He WD, conducted with the \textsc{Arepo} code.
With this simulation, we aim to provide accurate initial conditions of a merged hybrid HeCO WD that allow for a detailed long-term study of the nucleosynthesis in such objects and consequently determining if the predicted \ose{} ratios are in agreement with observed values.
We further investigate the long-term nucleosynthesis in different locations within the SoF, demonstrating the wide range of reaction regimes that are possible.
For the present work, we focus on studying the potential long-term evolution utilizing the conditions found immediately after the merger.
This exploratory study will guide future research that accurately accounts for the long-term evolution of the post-merger remnant object.

In Section\;\ref{sec:methods}, we describe the methods and codes used in our work.
Section\;\ref{sec:results} presents the results of our hydrodynamic simulation, nucleosynthesis results, and long-term evolution of the composition.
Lastly, we discuss our results in the context of other works in Section\;\ref{sec:discussion} before giving some concluding remarks in Section\;\ref{sec:conclusion}.

\section{Methods}\label{sec:methods}

\subsection{Hydrodynamic simulation}

We model the hydrodynamic evolution of the WD binary system in 3D using the code \textsc{Arepo} \citep{springel2010a,pakmor2011d,pakmor2016a,weinberger2020a}. \textsc{Arepo} solves the equations of ideal MHD using a second-order finite-volume approach on an unstructured moving Voronoi mesh. Explicit refinement and de-refinement conditions are employed for cells whose mass deviates by more than a factor of two away from the target mass of the cells ($M_\mathrm{target} = 5\times10^{-7}\,\msun$). Self-gravity is treated as Newtonian and the WD matter is modeled with a Helmholtz equation of state \citep{timmes2000a}, including Coulomb corrections \citep{yakovlev1989a}. 

We couple a $55$ isotope nuclear reaction network to the hydrodynamics, which includes the isotopes: n, p, $^4$He, $^{11}$B, $^{12-13}$C, $^{13-15}$N, $^{15-17}$O, $^{18}$F, $^{19-22}$Ne, $^{22-23}$Na, $^{23-26}$Mg, $^{25-27}$Al, $^{28-30}$Si, $^{29-31}$P, $^{31-33}$S, $^{33-35}$Cl, $^{36-39}$Ar, $^{39}$K, $^{40}$Ca, $^{43}$Sc, $^{44}$Ti, $^{47}$V, $^{48}$Cr, $^{51}$Mn, $^{52,56}$Fe, $^{55}$Co, and $^{56,58-59}$Ni. The reaction rates are provided by the JINA REACTLIB database \citep{cyburt2010a} and the network is active in all cells where  $T\,{>}\,2\,{\times}\,10^7\,\mathrm{K}$ and $\rho\,{>}\,1\,{\times}\,10^4\,\mathrm{g}\,\mathrm{cm}^{-3}$. 

The initial stellar binary system consists of a $0.6\,\msun$ CO WD ($50\,\%/50\,\%$ by mass) and a $0.3\,\msun$ He WD, both having an initial uniform temperature of $1\times10^6\,\mathrm{K}$.
Our simulation follows a setup similar to the one presented in \citet{moran-fraile2024a}. 
To ensure hydrostatic equilibrium in the stellar models, the WDs are relaxed in isolation for ten dynamical timescales of each star, following the procedure described in \citet{pakmor2012b,ohlmann2017a}. 

After the relaxation procedure, both WDs are placed together into a simulation box with size $7.7\times 10^{12}\,\mathrm{cm}$ and uniform background density of $1\times10^{-8}\,\mathrm{g}\,\mathrm{cm}^{-3}$.
To avoid strong tidal forces on the WDs as they are placed in the same box, the initial orbital separation is chosen to be $7.7\,\times10^9\,\mathrm{cm}$, which is equal to two times the separation at the point of Roche-lobe overflow (RLOF).
We artificially reduce the orbital separation following the inspiral procedure described by \citet{pakmor2021a} until reaching the point of RLOF, after which we let the system evolve freely. At this point we also enable the nuclear reaction network.
The simulation is followed for a total of $4000\,\mathrm{s}$ after RLOF.

\subsection{Nucleosynthesis}\label{sec:met_nucl}
We perform nucleosythesis calculations in postprocessing using the NuGrid\footnote{\url{nugridstars.org}} nuclear reaction network \citep{pignatari2016a,ritter2018c}, specifically the modified version described by \citet{jones2019a}. 
Here, we include a total of 5213 isotopes as listed in \citet[Table 1]{jones2019a}, in particular $\mathrm{n}$, \iso{H}{1-3}, \iso{He}{3-6}, \iso{Li}{7-9}, \iso{Be}{7-12}, \iso{B}{8-14}, \iso{C}{11-18}, \iso{N}{11-21}, \iso{O}{13-22}, \iso{F}{17-26}, \iso{Ne}{17-41}, \iso{Na}{19-44}, \iso{Mg}{20-47}, \iso{Al}{21-51}, and \iso{Si}{22-54}; although heavier isotopes are included, they are not relevant for the conditions studied here, and the effective network will only include up to around 300 isotopes.
For details on the included reaction rates and data sources, we refer the reader to \citet{jones2019a} and references therein.
We note that, in contrast to \citet{jones2019a}, we do not use the weak reaction rates of \citet{nabi2004a} as they are not relevant to our study.

To obtain detailed nucleosynthetic yields of the hydrodynamic simulation, we advect $10^6$ equal-by-mass Lagrangian tracer particles in the flow and record the temperature and density along their trajectories.
This data is then used as input for the NuGrid nuclear network calculations, from which we obtain detailed abundances for each particle at the end of the hydrodynamic simulation.

We further utilize the single-zone mode of the NuGrid network to compute the long-term nucleosynthesis in the SoF.
Here, we calculate radially averaged shells (temperature, density, and composition) of the post-processed tracer particles and use the resultant thermodynamic state and composition in each shell as the initial conditions for the single-zone models.
These radial zones are then extrapolated to $100\,\mathrm{years}$ assuming that the temperature and density remain constant for this duration.
We find that towards the end of the hydrodynamic simulation the temperature and density each approach a constant value in each shell; therefore we do not expect any further short-term changes in these quantities. 
However, without further long-term modeling efforts\footnote{See, e.g., \citet{schneider2020a} for long-term models of the merger product of massive stars}, it is impossible to accurately predict how the evolution of these shells (and the remnant in general) will proceed over longer timescales. We follow the approach already taken by \citet{staff2012a} and assume to conditions at the end of the hydrodynamic simulation to be maintained for an extended period of time, $100\,\mathrm{years}$ in our case.
As we are mainly interested in an exploratory study of the nucleosynthesis under the conditions found in the WD merger to establish potentially relevant reaction channels, rather than accurately modeling observed data, we deem this a sufficient approximation.

For the post-processing, we add an initial solar metallicity utilizing the data of \citet{asplund2009a}.
Here, we follow the scheme described by \citet{lach2020a} and convert CNO isotopes to \iso{N}{14} in the He WD and to \iso{Ne}{22} in the CO WD, as one would expect from the preceding stellar evolution.
In addition, we consider the presence of \iso{H}{1} in the He WD.
For the current study, we include this as a homogeneous mass fraction $X(^1\mathrm{H})=0.01$ throughout the WD (similar to \citealt{staff2012a}) as we are mostly interested in studying how the presence of free protons impacts the subsequent nucleosynthesis.
Although it is not entirely clear if a homogeneous H abundance is representative of the dynamic merger history, we deem this a sufficient approximation for this exploratory study. 
It is not yet clear to what extent mixing during the merger process could homogenize an initially stratified donor WD (e.g., a thin H shell on top of a He WD).
This will be further investigated in future studies.

\section{Results}\label{sec:results}
\subsection{Hydrodynamic merger simulation}\label{sec:res_hydro}
\begin{figure*}
    \centering
    \includegraphics[width=\textwidth]{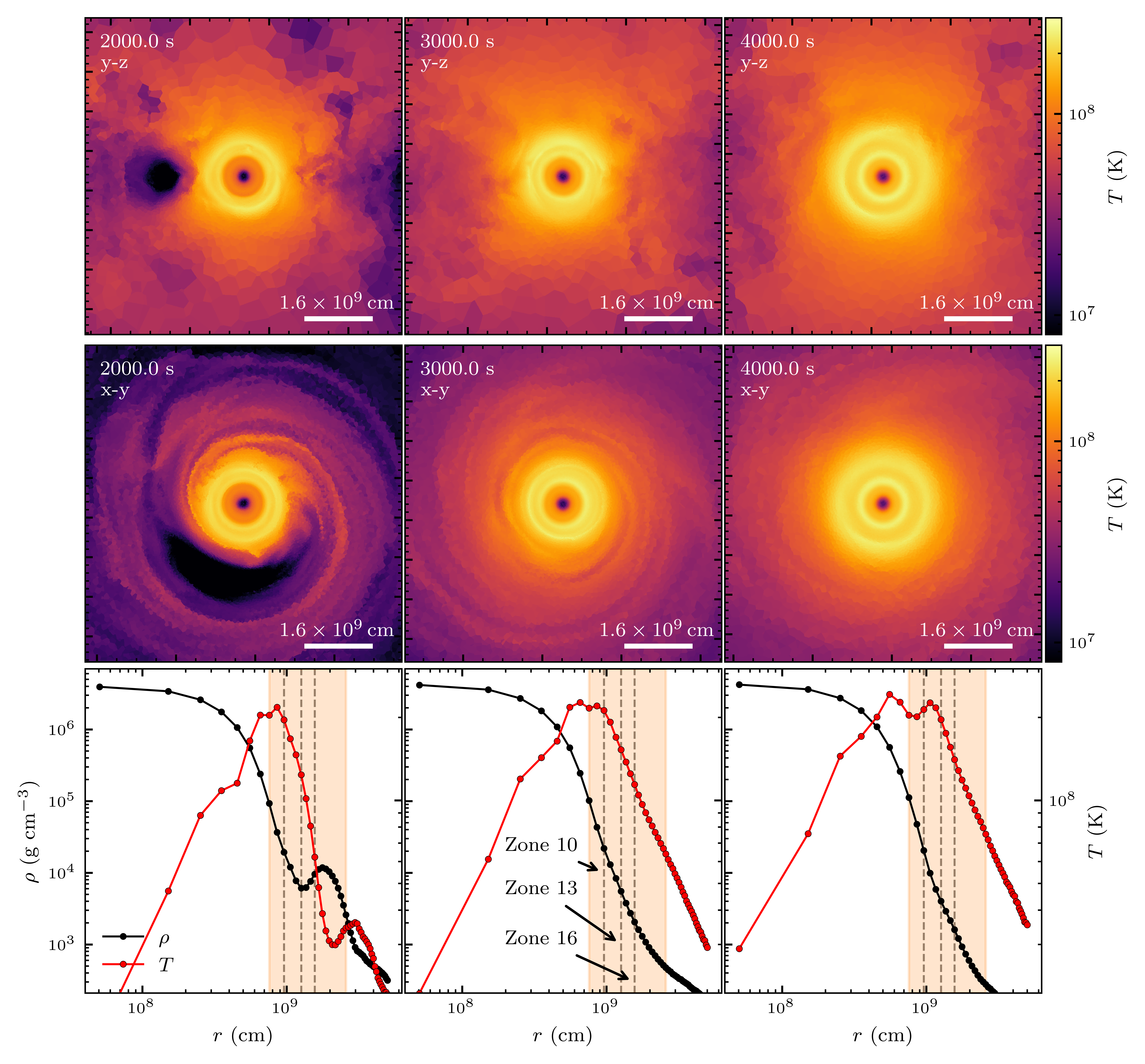}
    \caption{Illustration of the merged HeCO WD at different timesteps. The top and middle row show a planar temperature slice along the rotational and equatorial plane, respectively. Here, it can bee seen that initially strong asymmetries are present which get smoothed out over time. The lower row shows spherically averaged shells centered on the center of mass of the CO core. The orange band highlights the SoF investigated in the single-zone models; the dashed vertical lines indicate the zones discussed in-depth in Section\;\ref{sec:single_zone}.}
    \label{fig:merger}
\end{figure*}

Here, we only briefly outline the results of the merger simulation relevant for this work; a detailed analysis of this simulation will be given in its own dedicated study.
We start following the merger after the inspiral phase when RLOF sets in (we define this point in time as $t=0.0\,\mathrm{s}$), that is mass transfer from the He WD onto the surface of the CO WD begins. 
The evolution of the merger that follows is similar to the one outlined by, e.g., \citet{staff2012a,staff2018a,moran-fraile2024a,shiber2024a}.
The orbital separation decreases as a result of the mass transfer until the He WD is tidally disrupted and forms a shell around the CO WD, as well as an accretion disk; this process is illustrated in Figure~\ref{fig:merger}.
Initially, the He WD is not yet fully dissolved; see the cold blob in the $t=2000.0\,\mathrm{s}$ column in Figure\;\ref{fig:merger}, but eventually it forms a mostly spherical shell over the following $1000\,\mathrm{s}$ around the CO core\footnote{The merger also results in the formation of an accretion disk.
However, since no burning takes place in this region, we neglect it in the following discussion.}.
This shell continuously heats up, mostly because of the onset of nuclear burning, that is triple-$\alpha$ reactions.
The formation of this He burning layer, the SoF, is illustrated in Figure\;\ref{fig:merger}.
At $t=4000\,\mathrm{s}$ the shell has settled into a quasi-equilibrium state and He burning sustains the temperature at a constant value of a few $10^8\,\mathrm{K}$.
At this point,the density has also equilibrated into a steady state profile.

\subsection{Tracer particle yields}\label{sec:tppnp}
\begin{figure}
    \centering
    \includegraphics[width=\linewidth]{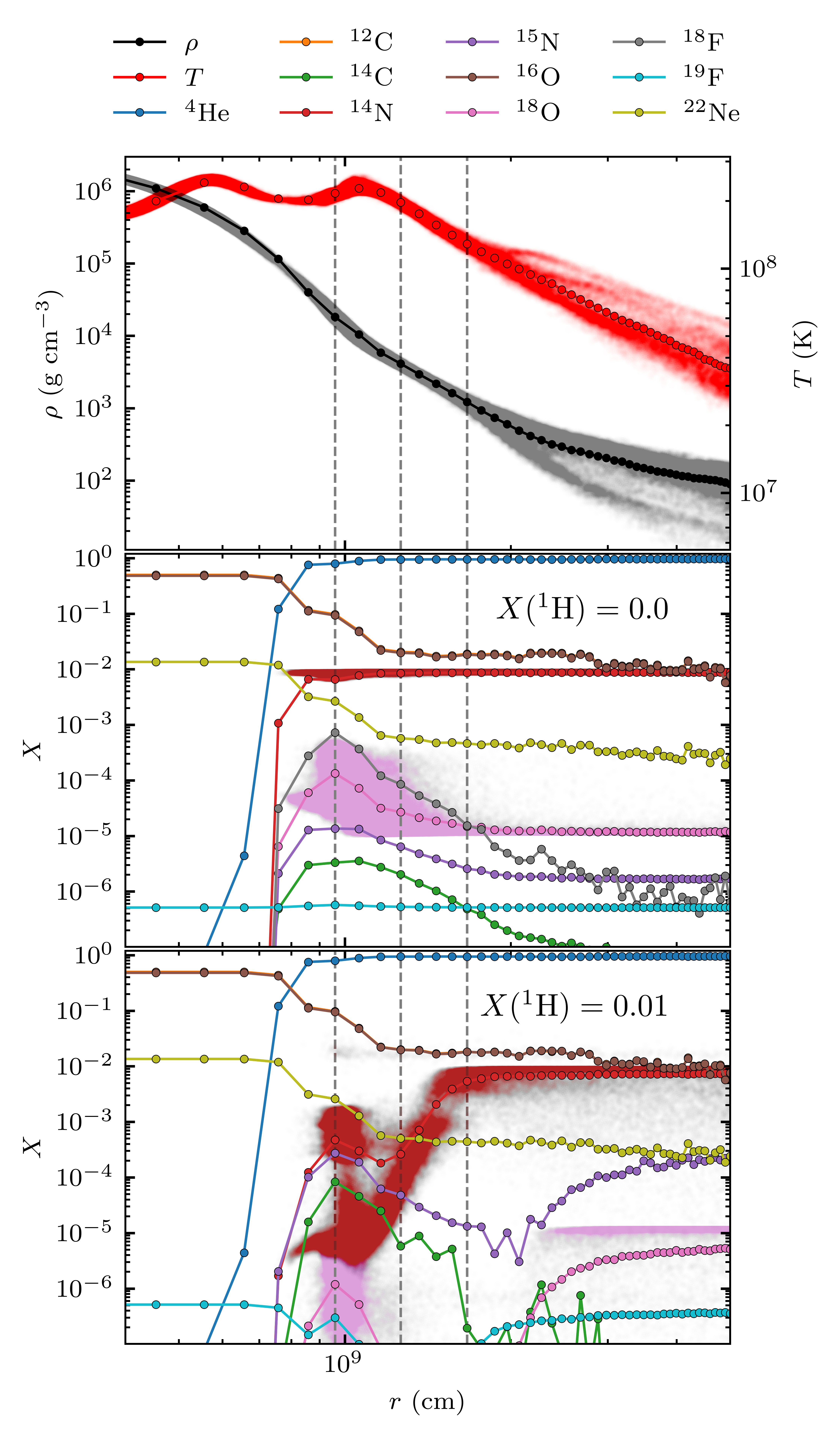}
    \caption{Spherically averaged tracer particle yields. Here, the binning is the same as in Figure\;\ref{fig:merger}. The top row shows the temperature and density values. The middle and bottom row illustrate the yields of various isotopes, both for the $X(^1\mathrm{H})=0.0$ and $X(^1\mathrm{H})=0.01$ case, respectively. In all rows, we also indicate the 3D nature of the tracer particles by adding each individual tracer particle value for select quantities. As in Figure;\ref{fig:merger}, the dashed vertical lines indicate the zones discussed in Section\;\ref{sec:single_zone}. Note that the abundances for \iso{C}{12} and \iso{O}{16} overlap in this figure.}
    \label{fig:yields}
\end{figure}
In Figure\;\ref{fig:yields}, we illustrate the radially binned, post-processed abundances of the tracer particles at the end of the hydrodynamic simulation.
Here, the burning takes place in an approximately $1.2\,{\times}\,10^9\,\mathrm{cm}$ wide shell above the CO core as suggested by the temperature profile already illustrated in Figure\;\ref{fig:merger}.
There are two immediate consequences:
First, any production of \iso{O}{18}, as well as any other isotopes of interest, takes place in a region of relatively high density and subsequently high optical depth.
This necessitates their transport into the photosphere by some mechanism, for example convection, in order to be observed.
Second, we also notice some dredge-up of \iso{O}{16} from the core into the He shell\footnote{\iso{C}{12} is dredged up in equal amounts, but we are mostly focused on \iso{O}{16} as this directly impacts the resultant \ose{} ratio.}, as already studied by \citet{staff2018a,shiber2024a}.
Within a $2\times10^8\,\mathrm{cm}$ thick layer above the CO core, we find substantial amounts of \iso{O}{16} originating from the core mixed into the He material.
The outer layers also show some contribution from dredge-up, but this depends much more strongly on the 3D structure of the merger; we leave a detailed analysis of this process to a dedicated study of the hydrodynamic simulation. 
We also highlight the radial zone shortly before the second temperature peak, that is, the innermost dashed vertical line in Figure\;\ref{fig:yields}.
Here, both temperature and density are sufficient for \iso{C}{12} and \iso{O}{16} to be produced; this shoulder in the \iso{C}{12} and \iso{O}{16} abundance does not originate exclusively from dredge up from the CO core during the merger, but has a clear contribution from nuclear burning.

In the $X(^1\mathrm{H}) = 0.0$ case, this region also produces most of the relevant isotopes such as \iso{F}{18} and \iso{O}{18}.
Moving to higher radii in this case is rather straightforward; burning becomes slower with decreasing temperatures and densities, until burning effectively ceases and the abundances approach their solar input value.

In contrast, the $X(^1\mathrm{H}) = 0.01$ case shows the clear impact of proton capture reactions such as
\begin{equation}
\begin{aligned}
    ^{18}\mathrm{O}(\mathrm{p},\gamma)^{19}\mathrm{F}(\mathrm{p},\gamma)^{20}\mathrm{Ne}\\
    ^{14}\mathrm{N}(\mathrm{p},\gamma)^{15}\mathrm{O},
\end{aligned}
\end{equation}
that destroy the initial solar abundance of key isotopes.
Only in the innermost burning zones is the timescale of reactions short enough to reproduce these isotopes close to the solar value.
These proton capture reactions are, in general, rather fast at the prevailing temperatures of around $2.2\times10^8\,\mathrm{K}$ (see \citealt[Figure 13]{staff2012a}) and occur on timescales of seconds.
The production of heavier isotopes, however, only takes place on the order of $100\,\mathrm{s}$.
The exceptions to this trend are \iso{C}{14} and \iso{N}{15}, which get overproduced compared to the $X(^1\mathrm{H}) = 0.0$ case through.
\begin{equation}
\begin{aligned}
    ^{18}\mathrm{O}(\mathrm{p},\alpha)^{15}\mathrm{N},\\   
    ^{15}\mathrm{N}(\mathrm{p},\gamma)^{16}\mathrm{O},\\
    ^{15}\mathrm{N}(\mathrm{n},\mathrm{p})^{14}\mathrm{C}.
\end{aligned}
\end{equation}

To illustrate the 3D nature of the merger simulation and the resulting burning products, we also show the individual tracer particle results for select data in Figure\;\ref{fig:yields}.
At the end of the hydrodynamic simulation, the spherical approximation appears to be sufficient for temperature and density.
Large deviations from this average only become noticeable outside the main burning region, visible, for example, in the lower density mode above $r=2\times10^9\,\mathrm{cm}$ in Figure\;\ref{fig:yields}\footnote{We note that in these outer layers, structures such as an accretion disc become relevant and a spherical model will no longer be a suitable approximation.}.
However, considering the burning products, the deviations from sphericity become more pronounced.
Looking, for example, at the \iso{N}{14} nucleosynthesis in the innermost burning shells, angular deviations in its abundance of up to three orders of magnitude can be observed.
This is an imprint of the dynamic merger history before temperature and density have stabilized; if \iso{N}{14} would have been synthesized after temperature and density have stabilized, one would expect a much more homogeneous distribution rather than the observed bimodal profile.
Consequently, the resultant long-term nucleosynthesis will also depend on the full 3D structure --- at the very least the variation in the initial \iso{N}{14} will critically determine the resultant \iso{O}{18} abundance --- and should be considered in future studies.

\subsection{Long term evolution}\label{sec:single_zone}
To establish the long-term evolution, we further process the tracer yields in spherically averaged single-zone models, as described in Section\;\ref{sec:met_nucl}.
The investigated region is illustrated in the second row of Figure\;\ref{fig:merger} in the highlighted orange band.
We assign each radial bin a zone number, starting from the innermost bin.
For example, the zones marked with dashed vertical lines in Figures\;\ref{fig:merger} and \ref{fig:yields} are considered zone $10$, $13$, and $16$, respectively.
Beyond the zones highlighted with the orange band, densities and temperatures become so low that nuclear burning is no longer relevant on the timescales studied here.
Although there are many reactions occurring under the conditions studied here, we focus our discussion only on reactions relevant to the production of \iso{O}{18}.
For a more extensive discussion in the context of RCB stars, see, e.g., \cite{crawford2020a} or for a more general discussion \cite{jose2020a}.

\begin{figure*}
    \centering
    \includegraphics[width=\textwidth]{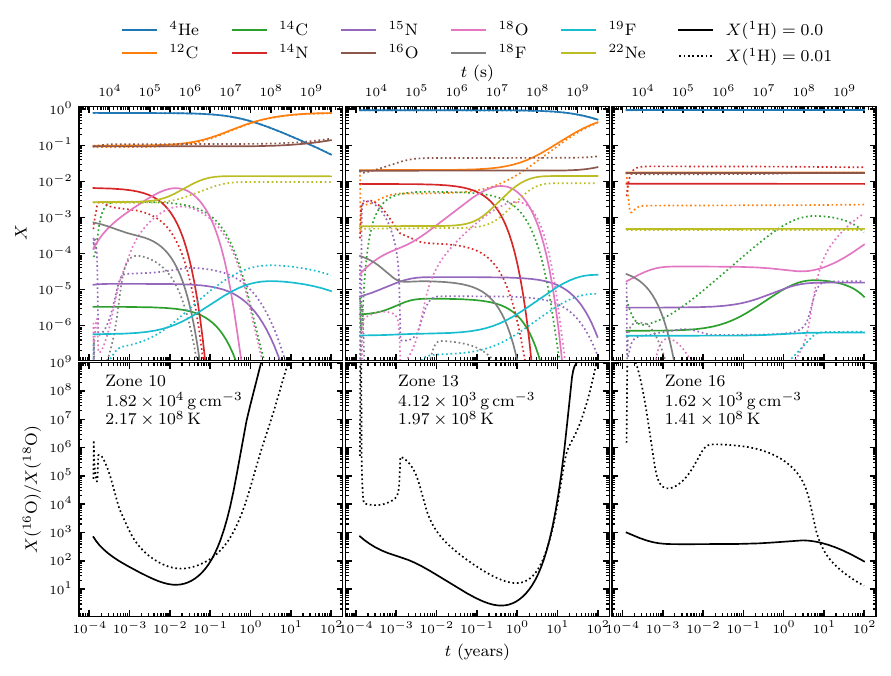}
    \caption{Illustration of the results of the single-zone network for three select zones. Top row: Abundance evolution of the zones discussed in Section\;\ref{sec:single_zone}. Here, we show both the $X(^1\mathrm{H})=0.0$ case (solid lines) and the $X(^1\mathrm{H})=0.01$ case (dotted lines). Bottom row: Evolution of the \ose{} ratio over time for both \iso{H}{1} fractions. In the individual columns, we also indicate the temperatures and densities present in the respective zones.}
    \label{fig:ppn}
\end{figure*}

The evolution of zones $10$, $13$, and $16$ is illustrated in Figure\;\ref{fig:ppn}, both for the case with and without initial \iso{H}{1}.
We also give the initial conditions for each zone in Table~\ref{tab:initial_abundances} for both temperature and density, as well as select isotopes with noticeable abundances in the respective case.
While the most immediate difference between the zones is the overall reaction timescale, the detailed reactions also differ.

\begin{table}
    \centering
    \caption{Initial conditions for the zones explored in Section\;\ref{sec:single_zone} for temperature and density, as well as mass fractions of select isotopes, for both the cases with and without an initial proton seed.}
    \label{tab:initial_abundances}
    \begin{tabular}{l|ccc}
     & Zone $10$ & Zone $13$ & Zone $16$ \\
    \hline
    $T$ ($10^8$ K) & 2.17 & 1.97 & 1.41 \\
    $\rho$ ($10^3$ g\,cm$^{-3}$) & 18.2 & 4.12 & 1.62 \\
    \hline
     & & $X(^1\mathrm{H}) = 0.0$ & \\
    $^4\mathrm{He}$ & 0.794 & 0.946 & 0.952 \\
    $^{12}\mathrm{C}$ & 0.098 & 0.021 & 0.018 \\
    $^{16}\mathrm{O}$ & 0.093 & 0.020 & 0.017 \\
    $^{14}\mathrm{N}$ & 0.007 & 0.009 & 0.009 \\
    $^{22}\mathrm{Ne}$ & 0.003 & 0.001 & 0.001 \\
    \hline
     & & $X(^1\mathrm{H}) = 0.01$ & \\
    $^4\mathrm{He}$ & 0.793 & 0.942 & 0.945 \\
    $^{12}\mathrm{C}$ & 0.097 & 0.020 & 0.017 \\
    $^{16}\mathrm{O}$ & 0.095 & 0.020 & 0.017 \\
    $^{14}\mathrm{N}$ & 0.000 & 0.000 & 0.004 \\
    $^{1}\mathrm{H}$ & 0.000 & 0.004 & 0.006 \\
    \end{tabular}
\end{table}

\subsubsection{Zone $10$}
In zone $10$ (left column in Table\;\ref{tab:initial_abundances}), in the $X(^1\mathrm{H})=0.0$ case, reactions occur relatively fast.
At early times
\begin{equation}
\begin{aligned}
    ^{14}\mathrm{N}(\alpha,\gamma)^{18}\mathrm{F}(\beta^+)^{18}\mathrm{O}&,\\
    ^{18}\mathrm{F}(\mathrm{n},\alpha)^{15}\mathrm{N}&,\\
    \rm{and}\quad^{18}\mathrm{O}(\mathrm{p},\alpha)^{15}\mathrm{N}& 
\end{aligned}
\end{equation}
are the most relevant reactions; the reactions that destroy \iso{F}{18} and \iso{O}{18} are, however, subdominant compared to the production reactions.
Apart from triple-$\alpha$ reactions, the largest mass flux is found in $^{14}\mathrm{N}(\alpha,\gamma)^{18}\mathrm{F}$, rather than, for instance, $^{12}\mathrm{C}(\alpha,\gamma)^{16}\mathrm{O}$.
We find that \iso{O}{18} is also produced by
\begin{equation}
    ^{14}\mathrm{C}(\alpha,\gamma)^{18}\mathrm{O}
\end{equation}
where \iso{C}{14} is produced through the neutron poisoning reaction $^{14}\mathrm{N}(\mathrm{n},\mathrm{p})^{14}\mathrm{C}$.
This reaction is completely subdominant compared to $^{19}\mathrm{F}(\beta^+)^{18}\mathrm{O}$, but is maintained for longer timescales when the \iso{N}{14} and subsequently \iso{F}{18} fuel is exhausted around $0.1\,\mathrm{years}$.
Neutrons are produced at a low rate through various reactions, such as
\begin{equation}
\begin{aligned}
    ^{13}\mathrm{C}(\alpha,\mathrm{n})^{16}\mathrm{O}&,\\
    ^{17}\mathrm{O}(\alpha,\mathrm{n})^{20}\mathrm{Ne}&,\\
    \rm{and}\quad^{18}\mathrm{O}(\alpha,\mathrm{n})^{21}\mathrm{Ne}.& 
\end{aligned}
\end{equation}

After $1\times10^{-3}\,\mathrm{years}$, the reaction
\begin{equation}
    ^{18}\mathrm{O}(\alpha,\gamma)^{22}\mathrm{Ne}
\end{equation}
becomes the dominant destruction channel for \iso{O}{18} until all available \iso{O}{18} is exhausted after about $1\,\mathrm{year}$.
Beyond this point in time, \iso{Ne}{22} is produced mainly by
\begin{equation}
    ^{19}\mathrm{F}(\alpha,\mathrm{p})^{22}\mathrm{Ne}
\end{equation}
whereby \iso{F}{19} has been steadily built up through
\begin{equation}
    ^{15}\mathrm{N}(\alpha,\gamma)^{19}\mathrm{F}
\end{equation}
before.
After a decade, \iso{O}{16} production starts to increase as well and \iso{He}{4} slowly depletes.

Considering the $X(^1\mathrm{H})=0.01$ case, the initial $1\times10^{-3}\,\mathrm{years}$ exhibit noticeably different reactions.
In addition to the previously mentioned ones, reactions such as
\begin{equation}
\begin{aligned}
    &^{18}\mathrm{F}(\mathrm{p},\alpha)^{15}\mathrm{O}\\
    \rm{and}\quad&^{17}\mathrm{O}(\mathrm{n},\alpha)^{14}\mathrm{C}
\end{aligned}
\end{equation}
are now rather efficient (due to the build up of \iso{O}{17} from $^{16}\mathrm{O}(\mathrm{p},\gamma)^{17}\mathrm{N}(\beta^+)^{17}\mathrm{O}$ and $^{16}\mathrm{O}(\mathrm{n},\gamma)^{17}\mathrm{O}$), destroying available \iso{F}{18} and building up \iso{C}{14}.
This initial burst of reactions involving protons and neutrons is clearly visible in the rapid abundance change in Figure \ref{fig:ppn} and the stark contrast to the $X(^1\mathrm{H}) = 0.0$ case is evident\footnote{We note that other reactions involving CNO isotopes are also effective during this period and noticeably change the composition going into the next phase.}.
After this initial burst, these reactions subside, and the evolution is similar to that discussed above, albeit from a different starting point.
However, now the production of \iso{O}{18} through $^{14}\mathrm{C}(\alpha,\gamma)^{18}\mathrm{O}$ is comparable to $^{18}\mathrm{F}(\beta^+)^{18}\mathrm{O}$ and accelerates its production.
Additionally, the reaction
\begin{equation}
    ^{18}\mathrm{F}(\mathrm{n},\mathrm{p})^{18}\mathrm{O}
\end{equation}
shows a substantial flux close to that of the beta decay reaction.
It can also be seen in Figure\;\ref{fig:ppn} that the production of \iso{O}{18} through \iso{C}{14} persists for much longer timescales even after \iso{F}{18} is mostly depleted.
After $1\,\mathrm{year}$, \iso{O}{18} has been burned to \iso{Ne}{22} and now \iso{F}{19} is the dominant fuel the production of \iso{Ne}{22}.

In this zone, we also observe a large initial \iso{C}{12} and \iso{O}{16} fraction, likely caused by dredge-up and C burning during the early stages of the merger.
For the proton-rich case, this is further increased by a small amount due to $^{13}\mathrm{C}(\alpha,\mathrm{n})^{16}\mathrm{O}$.

\subsubsection{Zone $13$}
Looking at zone $13$ (see the middle column of Table\;\ref{tab:initial_abundances}), starting with the case without additional \iso{H}{1}, we find that, apart from longer reaction timescales, the picture is overall similar to zone $10$.
\iso{O}{18} is produced from \iso{N}{14} until \iso{N}{14} is exhausted and eventually \iso{O}{18} is destroyed again by $\alpha$-captures to \iso{Ne}{22}.
However, there are some notable differences.
Importantly, within the first $1\times10^{-2}\,\mathrm{years}$ there is a strong production of \iso{N}{15} from \iso{F}{18}, \iso{O}{18}, and \iso{O}{15}.
In addition, substantial amounts of \iso{C}{14} are formed by $^{14}\mathrm{N}(\mathrm{n},\mathrm{p})^{14}\mathrm{C}$.
Generally, neutron and proton capture reactions also show comparably large fluxes, suggesting that a large initial abundance of neutrons and protons remains after the initial merger phase; indeed, the angle-averaged tracer abundances show an increase in the proton mass fraction by up to two orders of magnitude in zone $13$ compared to zone $10$.
This is mostly due to the reduced abundance of \iso{C}{12} and \iso{O}{16} in zone $13$ compared to that in zone $10$, which would otherwise dominate n- and p-capture reaction, thereby absorbing most of the free n and p.

After around $1\times10^{-2}\,\mathrm{years}$, most of these reactions stop, and the largest fluxes are in the \iso{O}{18} and \iso{Ne}{22} producing reactions.
Here, \iso{O}{18} production accelerates as it (and \iso{F}{18}) is no longer destroyed by, e.g. $^{18}\mathrm{O}(\mathrm{p},\alpha)^{15}\mathrm{N}$.
Similarly to zone $10$, \iso{F}{18} eventually depletes and \iso{O}{18} is only produced (at a comparably small rate) from \iso{C}{14} until its exhaustion.
The \iso{Ne}{22} production also slows down significantly once \iso{O}{18} is burned away.

Moving to the case with $X(^1\mathrm{H}) = 0.01$, we see an even larger impact of reactions involving protons than in zone $10$.
In the first few $10^{-5}\,\mathrm{years}$, we observe an intense burst of cold CNO cycling, particularly the first cold CNO cycle burning \iso{C}{12} to \iso{N}{14} and \iso{N}{15}.
After this phase, we still find many neutron and proton capture reactions, but at a lower rate.
Although severely depleted by $^{14}\mathrm{C}(\mathrm{p},\gamma)^{15}\mathrm{N}$, \iso{C}{14} is now produced in large amounts by $^{17}\mathrm{O}(\mathrm{n}, \alpha)^{14}\mathrm{C}$, $^{14}\mathrm{N}(\mathrm{n},\mathrm{p})^{14}\mathrm{C}$, and to a lesser extent by $^{13}\mathrm{C}(\mathrm{n},\gamma)^{14}\mathrm{C}$.
\iso{O}{18} is produced in small amounts by residual \iso{F}{18}, which in turn is produced slowly from \iso{N}{14}.
Importantly, the $^{18}\mathrm{F}(\mathrm{n},\alpha)^{15}\mathrm{N}$ flux is comparable to the $^{14}\mathrm{N}(\alpha,\gamma)^{18}\mathrm{F}$ flux and thus there is barely any buildup of \iso{F}{18}.
Until around $1\times10^{-2}\,\mathrm{years}$ $^{15}\mathrm{N}(\mathrm{p},\alpha)^{12}\mathrm{C}$ contributes in large parts to the production of \iso{C}{12}; afterwards, its production is dominated again by triple-$\alpha$ reactions.
During this phase, large amounts of \iso{O}{16} are produced as well by $^{13}\mathrm{C}(\alpha,\mathrm{n})^{16}\mathrm{O}$.

At $1\times10^{-3}\,\mathrm{years}$, $^{14}\mathrm{C}(\alpha,\gamma)^{18}\mathrm{O}$ becomes the dominant \iso{O}{18} production pathway, which can be seen in Figure\;\ref{fig:ppn} by the resurgence of \iso{O}{18} that has been previously destroyed by $^{18}\mathrm{O}(\mathrm{p},\alpha)^{15}\mathrm{N}$.
This is in so far interesting as previous work, e.g., \cite{staff2012a,crawford2020a}, does not consider \iso{C}{14} and thus does not observe this second resurgence in \iso{O}{18}.

As described before, these processes eventually produce \iso{Ne}{22} until \iso{O}{18} is completely depleted.

\subsubsection{Zone $16$}

In zone $16$ (see the right column of Table\;\ref{tab:initial_abundances}), the picture changes again, not only through longer reaction timescales, but also the possible reactions at the comparably low densities and temperatures found here.

Without additional protons, this zone initially mostly exhibits $^{18}\mathrm{F}(\beta^+)^{18}\mathrm{O}$ with $^{14}\mathrm{N}(\alpha,\gamma)^{18}\mathrm{F}$ slowly beginning to produce new \iso{F}{18} until around $1\times10^{-3}\,\mathrm{years}$.
This reaction is, however, relatively slow compared to the production of \iso{N}{14} and does not significantly impact the \iso{N}{14} abundance.
In this case the reaction $^{18}\mathrm{O}(\mathrm{p},\alpha)^{15}\mathrm{N}$ is also relatively efficient.

After approximately $1\times10^{-3}\,\mathrm{years}$, \iso{O}{18} is produced at the same rate as it is destroyed.
Later, at $1\times10^{-2}\,\mathrm{years}$, \iso{C}{14} is now produced at a noticeable rate, further boosting \iso{O}{18} production.
Only after one year does the \iso{O}{18} abundance decrease through $^{18}\mathrm{O}(\mathrm{p},\alpha)^{15}\mathrm{N}$ (instead of $\alpha$-captures on \iso{Ne}{22}) while \iso{C}{14} gets slowly burnt and the \iso{C}{14} production stops.
Overall, on the timescales simulated here, \iso{Ne}{22} is barely produced as the conditions are insufficient.

In zone $16$, the impact of initial protons, i.e. the $X(^1\mathrm{H}) = 0.01$ case, is similar to zone $13$.
An initial burst of cold CNO cycling destroys \iso{C}{12} and \iso{N}{15}.
The proton rich environment enables significant production of \iso{N}{14} (a notable difference to zone $13$ where the \iso{N}{14} amount produced does not exceed the initial abundance and remains below the $X(^1\mathrm{H}) = 0.0$ value), which eventually will lead to an overall higher \iso{O}{18} production.
For the most part, the remaining reaction pathways are comparable to the ones discussed for zone $13$ with an initial destruction of \iso{O}{18} and its production from \iso{F}{18} and \iso{C}{14}.

Regardless of the additional proton seed, \iso{Ne}{22} is not produced in this case either, at least not on the timescales considered here.

\subsubsection{$^{16}\mathrm{O}/^{18}\mathrm{O}$ ratio}
\begin{figure*}
    \centering
    \includegraphics[width=\textwidth]{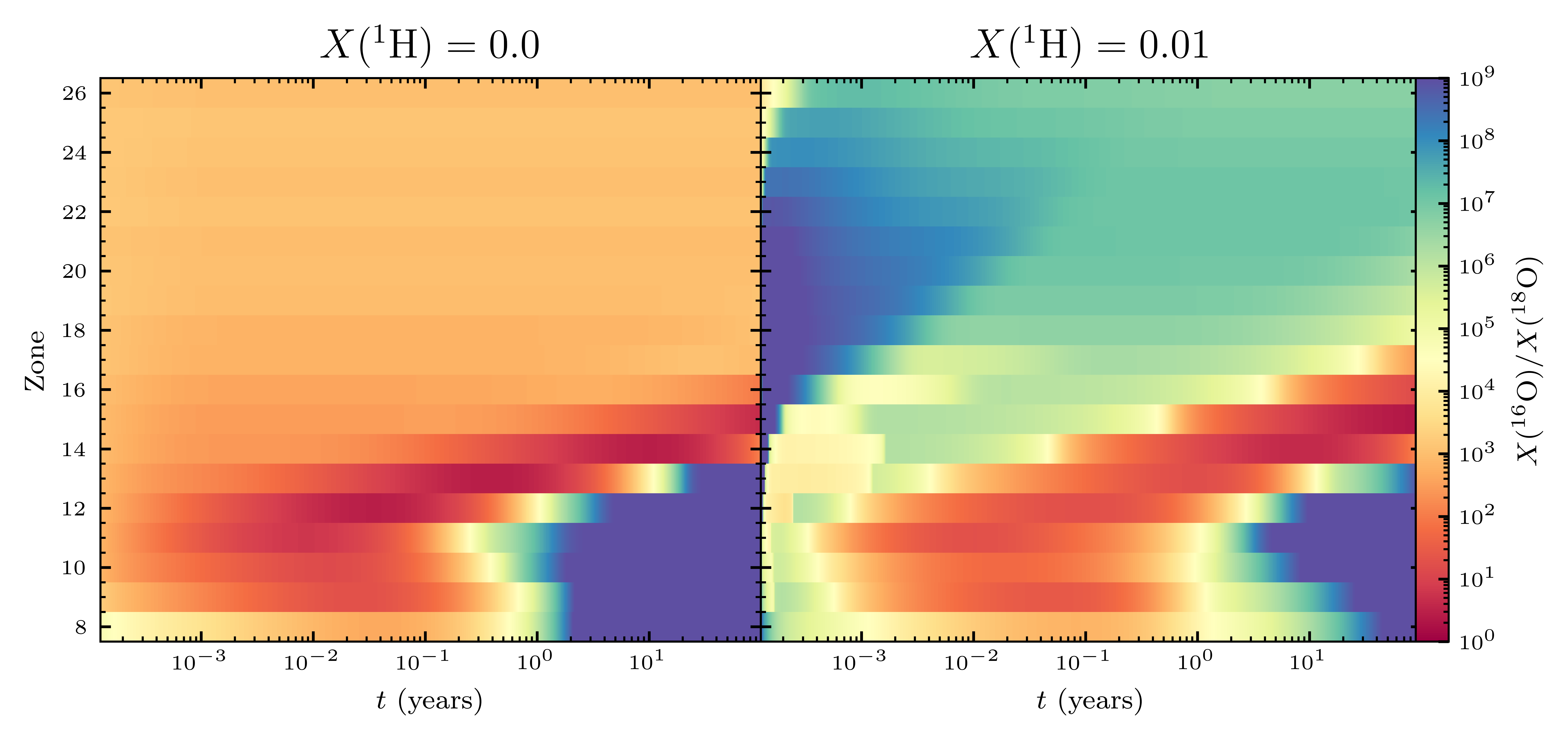}
    \caption{Illustration of the time evolution of the \ose{} ratio for each zone in the region highlighted in Figure\;\ref{fig:merger}. The left column shows the $X(^1\mathrm{H}) = 0.0$ case and the right column the $X(^1\mathrm{H}) = 0.01$ case.}
    \label{fig:zones}
\end{figure*}
Lastly, we present the evolution of the \ose{} ratios over time in our model.
Figure\;\ref{fig:ppn} shows the ratios for the zones discussed in the previous sections and Figure\;\ref{fig:zones} shows the \ose{} ratios for all $19$ zones highlighted in Figure\;\ref{fig:merger}.

Focusing first on Figure\;\ref{fig:ppn}, we find that in zone $10$ a minimal value of 10 is reached; adding an initial fraction of protons increases this ratio by almost an order of magnitude due to the destruction of \iso{O}{18} through $^{18}\mathrm{O}(\mathrm{p},\alpha)^{15}\mathrm{N}$.
In both cases, this minimum is only maintained for a very short period of time, on the order of a few days to weeks after the merger.
Due to the high temperature and density, \iso{O}{18} is rapidly burned to \iso{Ne}{22}.
It should be noted that due to the dredge up of \iso{O}{16} discussed in Section\;\ref{sec:tppnp}, the \ose{} ratio in this case is increased by this enrichment with \iso{O}{16}.

Zone $13$ achieves a \ose{} ratio close to unity which would be in the range suggested by \citet{mehla2025a} to match observed RCB stars. 
Similar to zone $10$, the $X(^1\mathrm{H})=0.01$ case does not reach the same low ratio, and for both cases this phase does not last very long, at most a few years.

However, in zone $16$, the picture is slightly different.
First, reactions seem to be slow enough that a long phase of low \ose{} --- on the order of centuries --- could reasonably be sustained.
It is somewhat difficult to extrapolate, but if the temperature and density are maintained for long enough, the values $\lesssim 10$ seem to be achievable for both $X(^1\mathrm{H})$ cases.
Second, in this zone, the proton rich case reaches lower \ose{} ratios, at least temporarily.
In contrast to zone $13$, where \iso{O}{16} is produced in the initial phase due to $^{13}\mathrm{C}(\alpha,\mathrm{n})^{16}\mathrm{O}$, in zone 16 a small amount is destroyed by proton captures $^{16}\mathrm{O}(\mathrm{p},\gamma)^{17}\mathrm{F}$.
Third, and most importantly, in zone $16$ the conditions seem insufficient to rapidly burn \iso{O}{18} to \iso{Ne}{22}, suggesting that these outer zones are the most favorable for reaching RCB-like \ose{} ratios on timescales of $10^2-10^3\,\mathrm{years}$.

This brings us to Figure\;\ref{fig:zones}, where this clear trend towards outer zones is clearly evident.
Here, it can be seen that the inner zones are clearly unfavorable for reaching low \ose{} ratios.
Contamination of \iso{O}{16}, either from dredge up or produced during the merger, negatively impacts the ratios in this regime.
Moreover, the temperatures and densities cause any \iso{O}{18} to burn to \iso{Ne}{22} too quickly.
The further out one goes, the lower the achieved \ose{} ratios become (although the lower limit seems to be on the order of unity) and the production timescales seem to be increased.
In general, the addition of protons appears to increase the overall value of the \ose{} ratio, before \iso{O}{18} production from \iso{C}{14} sets in.
In contrast, in the outer zones $\gtrsim 15$ the initial \iso{H}{1} fraction leads to lower \ose{} ratios on the timescales studied here.

Although zones out to zone $16$ (on the timescales simulated here) seem to allow for \ose{} ratios compatible with observed RCB stars, we find that there is a wide range of timescales on which suitable values are reached.
The inner zones are clearly too hot to prevent \iso{O}{18} burning to \iso{Ne}{22}, while the outer layers may not produce it fast enough to explain observed \ose{} values.
The intermediate zones seem the most likely to produce the correct amounts of \iso{O}{18} and keep it from burning away for long enough.

\section{Discussion}\label{sec:discussion}
As can be expected from a merger of two WDs, the process is highly asymmetric and cannot easily be represented in a 1D model.
Although Figure\;\ref{fig:merger} suggests the resultant temperature and density distribution is well approximated by a spherical average (at least up to $0.74\,\msun$ of the total mass), the nucleosynthetic post-processing discussed in Section\;\ref{sec:tppnp} reveals that the dynamic history is clearly imprinted on the abundance of key species, importantly \iso{N}{14}.
Without further mixing, this will directly impact the resultant yields of \iso{O}{18}, particularly without additional protons that enable the production of new \iso{N}{14}.
Recent works such as \cite{shiber2024a,staff2018a} have investigated the multi-dimensional properties of the merger process and merged hybrid WD; however, they do not present detailed nucleosynthesis results, but rather focus on the impact of dredge-up.
\cite{longland2011a,longland2012a} employ tracer particles similar to our approach, but do not examine isotopic yields with respect to their asymmetry.
Studies that discuss detailed reaction pathways (e.g., \citealt{staff2012a,menon2013a,schwab2019a,crawford2020a}) only consider a 1D initial abundance distribution.
For practical reasons, this is a sufficient approximation, but neglects the fact that the initial \iso{N}{14} can vary by several orders of magnitude if protons are available.
For example, our zone $10$ has an initial \iso{N}{14} mass fraction of a few $10^{-3}$ while Figure\;\ref{fig:merger} shows that initial abundances as low as $10^{-5}$ are possible; the resultant \iso{O}{18} will likely also vary by a similar amount.
\cite{yoon2007a} and \cite{longland2011a} suggest that the entire burning shell will quickly become convective; however, while we find a similar thermal timescale $\tau_\mathrm{th} \approx 10^9\,\mathrm{s}$, we observe a nuclear timescale of $\tau_\mathrm{nuc} \approx 10^4\,\mathrm{s}$, much longer than their estimated $100\,\mathrm{s}$.
Although these numbers alone do not allow for a definitive statement for the timescale on which convection will develop, they suggest that our model may take longer before it becomes convective, making the outcome more sensitive to the asymmetric distribution of the initial fuel composition.
For the purposes of this exploratory study, we deem the assumption of constant temperature and density sufficient, as we are mainly interested in establishing the possible range of relevant reaction regimes.
Future studies aiming at matching models to observed RCB stars will need to account for the long term evolution of the conditions in the SoF.

In contrast to recent studies such as the work of \cite{staff2012a} (see also \citealt{staff2018a,shiber2024a}), who observe only a very thin SoF with a width of $1-2\times10^8\,\mathrm{cm}$, our SoF\footnote{In this context we define the SoF as the region where nuclear burning occurs on the timescales of our hydrodynamic simulation.} is much thicker on the order of $1\times10^9\,\mathrm{cm}$.
A likely explanation for this is that since we consider the energy release from nuclear burning, a larger area is heated up, allowing for nuclear reactions to occur.
This also has the consequence that for our simulation a single temperature and density average for the entire SoF would not be a good approximation as in the case of \citet{staff2012a}.
We note that although we additionally consider the energy released from nuclear burning (compared to \citealt{staff2012a,staff2018a,shiber2024a} who neglect this), we find a peak temperature in the SoF of slightly above $2\times10^8\,\mathrm{K}$, compared to the $3\times10^8\,\mathrm{K}$ in their mass ratio $q=0.5$ scenario.

Overall, the thicker SoF has two main consequences for the \ose{} ratio.
On the one hand, compared to other studies, we observe long-term \iso{O}{18} formation in the outer layers and, particularly, its production in regions too cold and low density to burn to \iso{Ne}{22}.
On the other hand, the dredge-up of \iso{O}{16} into the accreted material is no longer as impactful to the \ose{} ratio.
We also observe some mixing of core material with the accreted shell (in a roughly $10^8\,\mathrm{cm}$ thick region), but this layer is not critically relevant to what is likely the observable surface of the merged object.
Both of these consequences are favorable for achieving a low \ose{} ratio for the expected life-time of an RCB star.
However, this picture may change if convection sets in (or any other mechanism that changes temperature and density); depending on the thickness of the convecting layer, parts of the SoF, if not the entire shell, may become thoroughly mixed.
We leave a study of the impact of convection for future work.

Another indirect consequence of the thicker SoF is the emergence of $^{14}\mathrm{C}(\alpha,\gamma)^{18}\mathrm{O}$ as an additional, potentially dominant, production channel for \iso{O}{18}.
Other works (e.g. \citealt{staff2012a,crawford2020a}) neglect \iso{C}{14} in their networks --- given the conditions considered by them, rightfully so --- and thus overlook its contribution.
However, particularly in the outer shells, where a favorable \ose{} ratio is reached on long timescales and $^{18}\mathrm{O}(\alpha,\gamma)^{22}\mathrm{Ne}$ is suppressed, \iso{C}{14} becomes an important source of \iso{O}{18} and should be considered.

In our study, we also find that an initial source of protons can have a drastic impact on subsequent nucleosynthesis because it allows, for example, efficient CNO cycling and \iso{N}{14} production.
Although other works find similar abundance fractions for the He WD as the $X(^1\mathrm{H})=0.01$ explored by us (e.g., $1.51\%$ of \citealt{staff2012a}), it is not clear if a homogeneous distribution of protons is found in nature (e.g., \citealt{cunningham2020a}).
To what extent this surface layer will be mixed into the SoF during the merger remains to be seen and will be considered in future studies.
Furthermore, primordial \iso{He}{3} \citep{vangioni-flam2003a} could equally provide protons to the fuel mixture, ultimately allowing for enhanced \iso{N}{14} and subsequent \iso{O}{18} production.

Overall, we find that the \iso{O}{18} production is highly localized.
Factors such as the fuel distribution after the merger phase, varying thermodynamic conditions and consequentially different production channels contribute to a spatially and temporally varying \ose{} ratio.
In our scenario, although observed, dredge-up of \iso{O}{16} does not negatively impact the resultant \ose{} ratios, except in a few innermost shells, in which the temperatures are too high to maintain \iso{O}{18} for long periods of time in any case.
For several shells, we find \ose{} ratios that fall within the range of what is observed, that is, around $10^0-10^2$ \citep{mehla2025a}.
However, our study neglects the long-term effects of, for example, convection and mass loss over time.
Therefore, it remains to be seen whether the static stratification studied here is representative of nature.

Lastly, it is also not entirely clear which layers will contribute to observed spectra.
Although it is likely that the outer low \ose{} layers will be closer to the photosphere, we refrain from definitive statements and leave an investigation of this for future work.

\section{Conclusion}\label{sec:conclusion}
In this work, we examined the localized \iso{O}{18} production in a merger of a $0.6\,\msun$ CO WD and a $0.3\,\msun$ He WD.
In Section\;\ref{sec:res_hydro} we briefly outlined the merger and described the final merged HeCO WD structure, from which we derived 1D zones to study their long-term nucleosynthesis.
In Section\;\ref{sec:tppnp}, we highlighted the nucleosynthesis results following the hydrodynamic simulation.
Finally, in Section\;\ref{sec:single_zone}, we described the detailed long-term nucleosynthesis based on three characteristic zones in the SoF that illustrate the various burning regimes.
Here, we also discussed the spatially and temporally varying \ose{} ratios of our model.

In summary, we draw three main conclusions from this study.
First, we demonstrate that although the final merged HeCO WD at the end of our hydrodynamic simulation (at $4000\,\mathrm{s}$) is to good approximation spherically symmetric in the regions of interest, the dynamic history of the merger process gets imprinted in the composition of the SoF, particularly its inner layers, where nuclear reaction timescales are short.
This will directly impact subsequent nucleosynthesis unless convection or a similar process can erase the merger history on relatively short timescales.
Ultimately, this underscores the multidimensional nature of such mergers.

Second, our study reveals that the SoF can host a variety of different nucleosynthetic processes and so far unexplored reaction channels such as $^{14}\mathrm{C}(\alpha,\gamma)^{18}\mathrm{O}$ can become the dominant \iso{O}{18} production mechanism, or at least an important side channel.
Furthermore, nucleosynthesis is highly sensitive to the initial abundance of protons, and therefore future studies should take special care of isotopes such as \iso{H}{1} or \iso{He}{3}.

Third, and most importantly, we find that our model can produce \ose{} ratios well within the observed range and maintain these values for long periods of time.
Our results indicate a strong time and location dependence of their values.
In particular, we observe a much thicker SoF than in comparable previous studies, allowing for $^{14}\mathrm{C}(\alpha,\gamma)^{18}\mathrm{O}$  to become the dominant \iso{O}{18} production channel. Lower temperatures and densities also suppress $^{18}\mathrm{O}(\alpha,\gamma)^{22}\mathrm{Ne}$, therefore preventing the destruction of \iso{O}{18} over time.

In this study, we do not consider the impact of convection and other dynamic effects that can impact the stratification in the SoF.
This may be an important factor in determining the \ose{} ratio in the photosphere that ultimately can be observed.
Future studies should therefore focus on establishing to what extent such mechanisms impact the \ose{} ratio in the observable regions of the star.

\backmatter

\bmhead{Acknowledgements}
A.H., V.A., and M.V. are fellows of the International Max Planck Research School for Astronomy and Cosmic Physics at the University of Heidelberg (IMPRS-HD) and acknowledge financial support from IMPRS-HD.

This work was supported by the Klaus Tschira Foundation, the Deutsche Forschungsgemeinschaft (DFG, German Research Foundation) -- RO 3676/7-1, project number 537700965, and by the European Union (ERC, ExCEED, project number 101096243). Views and opinions expressed are, however, those of the authors only and do not necessarily reflect those of the European Union or the European Research Council Executive Agency. Neither the European Union nor the granting authority can be held responsible for them.

This work was supported by the U.S. Department of Energy through the Los Alamos National Laboratory. Los Alamos National Laboratory is operated by Triad National Security, LLC, for the National Nuclear Security Administration of U.S. Department of Energy (Contract No. 89233218CNA000001).



\section*{Statements and Declarations}

The authors have no competing interests to declare.





\end{document}